# Atomic-scale insights into electro-steric substitutional chemistry of cerium oxide

Haiwu Zhang, Ivano E. Castelli*, Simone Santucci, Simone Sanna, Nini Pryds, and Vincenzo Esposito*

Cerium oxide (ceria, $CeO_2$) is one of the most promising mixed ionic and electronic conducting materials. Previous atomistic analysis has covered widely the effects of substitution on oxygen vacancy migration. However, an in-depth analysis of the role of cation substitution beyond trivalent cations has rarely been explored. Here, we investigate soluble monovalent ($Li^+$, $Na^+$, $K^+$, $Rb^+$), divalent ($Fe^{2+}$, $Co^{2+}$, $Mn^{2+}$, $Mg^{2+}$, $Ni^{2+}$, $Zn^{2+}$, $Cd^{2+}$, $Ca^{2+}$, $Sr^{2+}$, $Ba^{2+}$), trivalent ($Al^{3+}$, $Fe^{3+}$, $Sc^{3+}$, $In^{3+}$, $Lu^{3+}$, $Yb^{3+}$, $Y^{3+}$, $Er^{3+}$, $Gd^{3+}$, $Eu^{3+}$, $Nd^{3+}$, $Pr^{3+}$, $La^{3+}$) and tetravalent ($Si^{4+}$, $Ge^{4+}$, $Ti^{4+}$, $Sn^{4+}$, $Hf^{4+}$, $Zr^{4+}$) cation substituents. By combining classical simulations and quantum mechanical calculations, we provide an insight into defect association energies between substituent cations and oxygen vacancies as well as their effects on the diffusion mechanisms. Our simulations indicate that oxygen ionic diffusivity of subvalent cation-substituted systems follows the order $Gd^{3+}$ > $Ca^{2+}$ > $Na^+$. With the same charge, a larger size mismatch with $Ce^{4+}$ cation yields a lower oxygen ionic diffusivity, i.e., $Na^+$ > $K^+$, $Ca^{2+}$ > $Ni^{2+}$, $Gd^{3+}$ > $Al^{3+}$. Based on these trends, we identify species that could tune the oxygen ionic diffusivity: we estimate that the optimum oxygen vacancy concentration ($x_{V_O^{\bullet\bullet}}$) for achieving fast oxygen ionic transport is 2.5% for $Gd_xCe_{1-x}O_{2-x/2}$, $Ca_xCe_{1-x}O_{2-x}$ and $Na_xCe_{1-x}O_{2-3x/2}$ at 800 K. Remarkably, such a concentration is not constant and shifts gradually to higher values as the temperature is increased. We find that co-substitutions can enhance the impact of the single substitutions beyond that expected by their simple addition. Furthermore, we identify preferential oxygen ion migration pathways, which illustrate the electro-steric effects of substituent cations in determining the energy barrier of oxygen ion migration. Such fundamental insights into the factors that govern the oxygen diffusion coefficient and migration energy would enable design criteria to be defined for tuning the ionic properties of the material, e.g., by co-doping.

## 1. Introduction

Materials based on cerium oxide, (ceria, $CeO_2$), have attracted an upsurge of interest due to their versatile nature for applications, as they display fast oxygen ion conductivity, efficient catalysis, and giant electro-chemo-mechanical response.[1-5]

Such fascinating effects originate from extrinsic doping, i.e., unique defect chemistry, where point defects, such as oxygen vacancies, are generated to balance the substitutional cations. Understanding the underlying mechanism that controls ionic transport and diffusion is of particular importance. Maximizing the oxygen ion conductivity of ceria is highly desirable for applications such as solid oxide electrolytes and oxide permeable membranes and has been intensively investigated over the last decades.[6-10] However, low conductivity is also relevant, since a substantial leakage current can lead to the easy breakdown of electromechanical devices, especially in the high electrical field range, e.g., > 50 kV cm$^{-1}$).[11-12]

The oxygen exchange and diffusion properties of $CeO_2$ are typically tuned by substituting $Ce^{4+}$ cations with trivalent, rare-earth cations to generate a high concentration of oxygen vacancies ($V_O^{\bullet\bullet}$).[13] Differences in the ionic radii, valence state, and electronic configuration of the substituent cations with $Ce^{4+}$, can induce repulsive elastic energy and attractive electronic energy between the oxygen vacancy and substituent cations. Such interactions give rise to oxygen vacancy–substituent cation associations, resulting in an ionic conductivity maximum with increasing doping concentration ('volcano-type' behavior).[14, 15] Extensive experimental and theoretical works have revealed that the ionic conductivity in $CeO_2$ increases with increasing substituent cation radius up to $Gd^{3+}$, $Eu^{3+}$, $Sm^{3+}$, and $Nd^{3+}$ but decreases afterwards.[16-17] Based on Density Functional Theories (DFT) simulations, Andersson *et al.*[18] proposed an improvement in the oxygen ion conductivity of $CeO_2$ by introducing a mixture of $Nd^{3+}$/$Sm^{3+}$ or $Pr^{3+}$/$Gd^{3+}$ to balance the attractive and repulsive interactions, thereby reducing the association energies. An alternative approach is to fabricate $CeO_2$ codoped with rare-earth ions ($Y^{3+}$, $Sm^{3+}$, $La^{3+}$) and alkaline earth ions ($Ca^{2+}$, $Sr^{2+}$) to tune the 'effective' ionic radius further and to improve the oxygen ion diffusivity. Pearce and Thangadurai found that $CeO_2$ ceramics co-doped with rare-earth ($La^{3+}$, $Sm^{3+}$) and alkaline earth ions ($Ca^{2+}$, $Sr^{2+}$) exhibit higher ionic conductivities than those seen in singly-doped systems.[19] According to Xu *et al.*,[20] the ionic conductivity of $Ce_{0.8+x}Y_{0.2-2x}Ca_xO_{1.9}$ (x=0-0.1) at 700 °C initially increased and then decreased as the $Ca^{2+}$ concentration was increased further (34.2, 47.2 and 19.2 mS cm$^{-1}$ for x=0, 0.05, and 0.1, respectively).

To date, a systematic and comparative investigation on monovalent ($M^+$), divalent ($M^{2+}$), trivalent ($M^{3+}$) and tetravalent ($M^{4+}$) cation-substituted systems is still missing, as investigations on tuning the ionic conductivity beyond trivalent, rare-earth ions-substituted $CeO_2$ are still rare.[21-23] In particular, the solution energy, the interaction between $M_{Ce}'''$, $M_{Ce}''$, $M_{Ce}^{\times}$ and $V_O^{\bullet\bullet}$ compared with that of $M_{Ce}'$ (Kröger-Vink notation[24] for defect centers $M^+$, $M^{2+}$, $M^{4+}$ and $M^{3+}$, respectively) and their effects on the oxygen ion diffusion dynamics and mechanisms are not well understood. Interpreting the nature of improved oxygen ion conductivity is further complicated by the fact that $CeO_2$ is often contaminated by Si, Al, Fe and Mn,[25, 26] which often segregate at surfaces and/or grain boundaries.[27, 28] Therefore, clarifying the influence of such 'unwanted' cations on the ionic conductivity of $CeO_2$-based materials is crucial. Furthermore, although an atomic-scale study on the defect chemistry of $M^+$, $M^{2+}$ and $M^{4+}$ substituted $CeO_2$ is still rare, promising phenomena have been observed in such systems, such as the improved redox properties of $CeO_2$-$ZrO_2$ by sodium ($Na^+$) inclusion,[29] the anomalously large dielectric properties in Ca:$CeO_2$ ceramics,[30] the improved $O_2$ uptake in $Cu^{2+}$-doped Pr:$CeO_2$,[31] and the enhanced oxygen storage capacity of $Ce_{1-x}Ti_xO_2$.[32]

Herein, we directly investigate the effects of $M^+$, $M^{2+}$, $M^{3+}$, and $M^{4+}$ cations on oxygen ion migration in $CeO_2$ through atomistic simulations, at the classical and quantum mechanical levels. We first investigated the solution energy of $M^+$, $M^{2+}$, $M^{3+}$, and $M^{4+}$ cations in $CeO_2$ and the association energy between the substituent cation and $V_O^{\bullet\bullet}$ by molecular statics (MS) lattice simulations. Using molecular dynamics (MD) simulations, we calculated the oxygen tracer diffusion coefficient ($D_O^*$) and the migration energy ($E_{mig}$). Finally, we calculated oxygen migration with various diffusion pathways by a detailed Nudged-Elastic-Band (NEB) method,[33] in the framework of DFT.

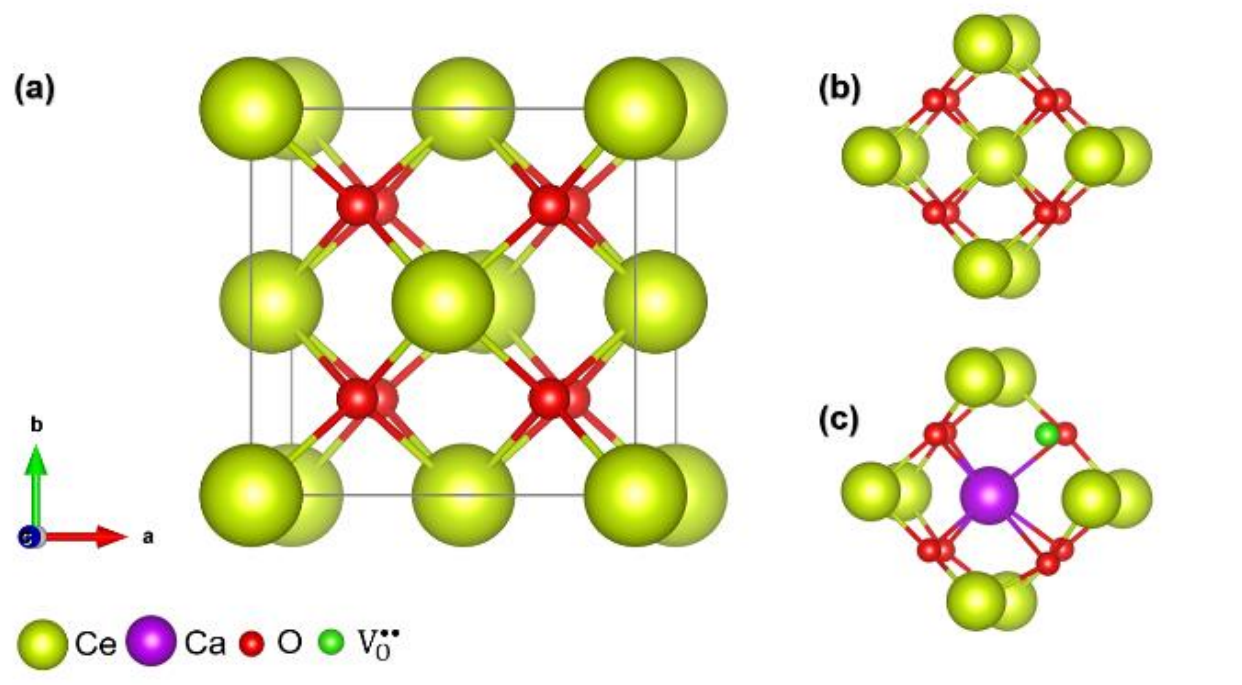

**Fig. 1** (a) Cubic fluorite-structured lattice of $CeO_2$; (b) local coordination of $Ce^{4+}$ cation; (c) local structure of one $Ca^{2+}$ cation with an oxygen vacancy at the nearest-neighboring site. Colors: $Ce^{4+}$, yellow; dopant cation, purple; oxygen, red; oxygen vacancy, green.

## 2. Computational methods

**Classical simulations**. Here, the interatomic forces are described using long-range Coulombic forces and Buckingham pair-potentials:

$$V_{ij} = \frac{Z_i Z_j}{4\pi\varepsilon_0 r_{ij}} + A_{ij}\exp\left(-\frac{r_{ij}}{\rho_{ij}}\right) - \frac{C_{ij}}{r_{ij}^6} \quad (1)$$

where $r_{ij}$ is the distance between ions $i$ and $j$, $Z_{i(j)}$ are the ion's valences, $\varepsilon_0$ is the permittivity of free space, and the parameters $A_{ij}$, $\rho_{ij}$ and $C_{ij}$ are the empirical Buckingham pair-potentials (the list of parameters is reported in the Supplementary Information, Table S1). The ionic polarizability of the ions is described using the shell model, and a harmonic spring is used to attach the massless shell with the ionic core.[34]

The defect energies are calculated by static lattice simulations using the Mott–Littleton approach,[35] as implemented within the GULP code.[36] We can thus summarize the defect energy ($E_{def}$) with the following equation:[36]

$$E_{def} = E_I(x) + E_{IIA}(x,y) + E_{IIB}(y) \quad (2)$$

where ionic coordinates and dipole moments determine $E_I$; $E_{IIA}$ is the interaction between region I and region II (interfacial region); and $E_{IIB}$ is determined by the displacements ($y$) in region IIA. Here we set the radii for regions I and II at 11 and 22 Å, respectively.

We calculated the binding energies ($E_{bind}$) of defect clusters, comprised of aliovalent/isovalent cations and oxygen vacancies, using the general relation:

$$E_{bind} = E_{associate} - (E_{isolated\ cation} + E_{isolated\ vacancy}) \quad (3)$$

where $E_{associate}$, $E_{isolated\ cation}$ and $E_{isolated\ vacancy}$ are the defect energy values of the defect cluster, the isolated aliovalent/isovalent cations and the isolated oxygen vacancy, respectively. A negative $E_{bind}$ indicates an attractive interaction between the substituent cations and the oxygen vacancy. We report the details for the defect reaction and the calculation of the solution energy ($E_{sol}$) in Supplementary Information (section 3).

All ions were treated as rigid ions (omitting the shell model) for the MD simulations. We controlled the temperature by a Nosé-Hoover thermostat with a significant time scale on which $P/T$ is relaxed (0.5 and 0.2 for $P_{damp}$ and $T_{damp}$, respectively), as implemented in the LAMMPS package.[37–39] We introduced the oxygen vacancies by randomly removing oxygen ions and compensated for them by lowering the charge of all $Ce^{4+}$ to 3.96 for the 'pure' system ($Ce^{3.96+}_{4000}O^{2-}_{7920}$) and for the $M^{4+}$-substituted systems ($Ce^{3.96+}_{3840}M^{4+}_{160}O^{2-}_{7920}$), or by acceptor cations, i.e., $M^+$ ($Ce^{4+}_{4000-x}M^{+}_{x}O^{2-}_{8000-3x/2}$), $M^{2+}$ ($Ce^{4+}_{4000-x}M^{2+}_{x}O^{2-}_{8000-x}$), and $M^{3+}$ ($Ce^{4+}_{4000-x}M^{3+}_{x}O^{2-}_{8000-x/2}$). At each temperature, the MD simulations were first equilibrated for at least 600 ps at constant pressure (NPT ensemble), followed by a production run for 600 ps in a NVT ensemble.

**Quantum mechanical calculations**. DFT calculations were performed using the Quantum Espresso code.[40] The general gradient approximation (GGA) was used according to the Perdew–Burke–Ernzerhof (PBE) functional to describe the exchange-correlation interaction (the Standard Solid State Pseudopotential, SSSP precision libraries).[41] We applied a Hubbard $U$ parameter of 5.0 eV on cerium, as recommended by Castleton *et al*.[42] We achieved convergence of the geometry and lattice energy with the cut-off energy of the plane-wave basis set and electronic density being 100 and 800 Ry, respectively. A comparison of the lattice parameters calculated with various methods is reported in Table S2. A 2×2×2 supercell of the fluorite structure (96 atoms) was used, with structure optimization using a 2 × 2 × 2 Monkhorst–Pack k-point grid. For the defect calculations, relaxation of the ion positions was conducted until the Hellmann–Feynman forces were lower than 1 meV Å$^{-1}$. The NEB method[33] with at least three images per calculation was used to obtain energy barriers and reaction paths (gamma point only).

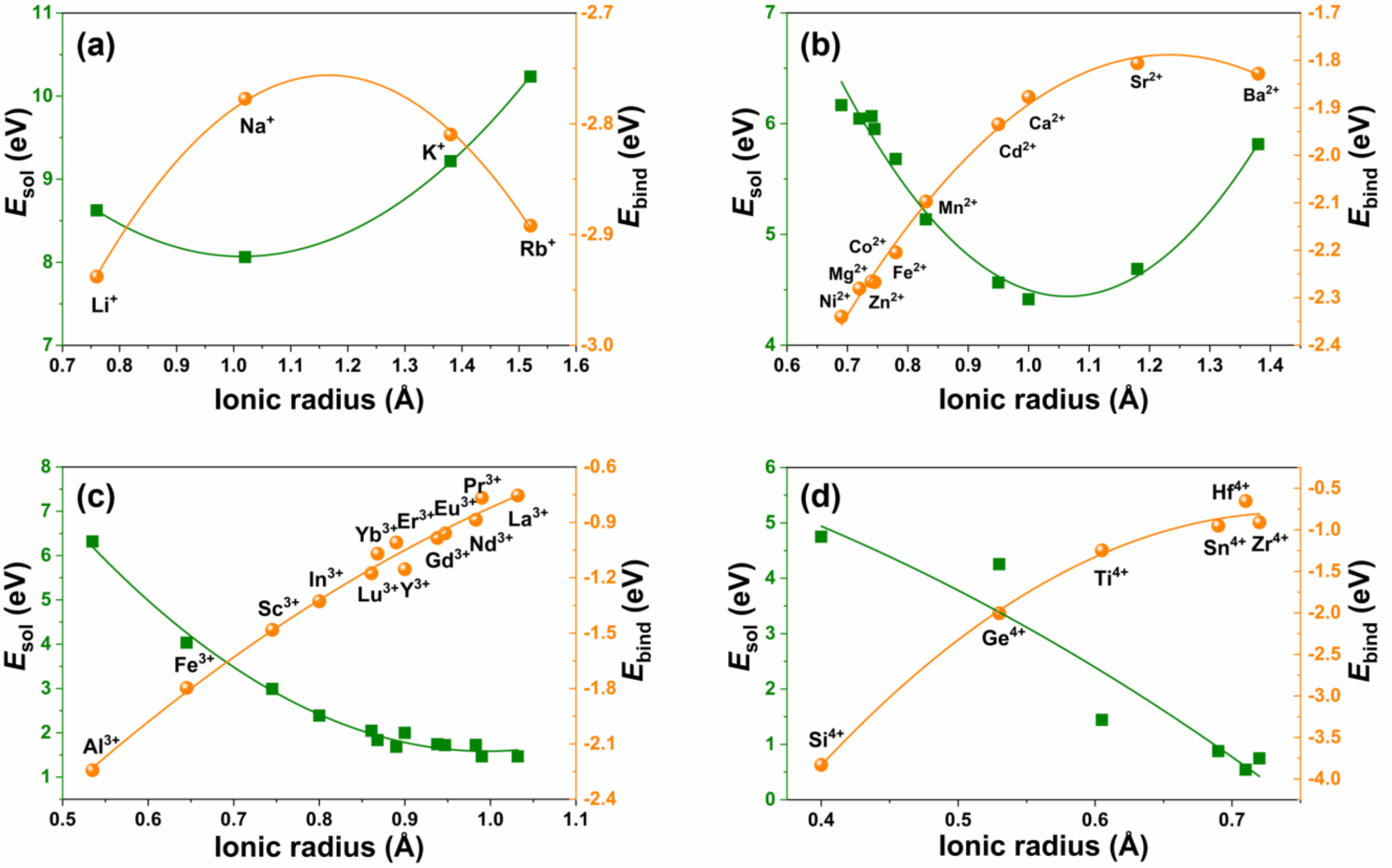

**Fig. 2** Solution energy and binding energy of (a) $M^{+}$; (b) $M^{2+}$; (c) $M^{3+}$ and (d) $M^{4+}$ cations in $CeO_2$ calculated by MS simulations. Since the ionic radii for eight-coordination values are not available for all of the ions considered, the values for six-coordination from Shannon[43] are used. For cations with a magnetic moment, the values for high-spin ionic radius are used.[43] The lines are used to guide the eyes only.

## 3. Results and discussion

To have a clear picture of the oxygen ion migration in various substituted $CeO_2$ instances, we investigate the key factors governing the association energy of oxygen vacancies with the cation defects incorporated. Firstly, we examined the trend of ion size for solution energy and association energy between the substituent cations and oxygen vacancies. Secondly, we examined the oxygen diffusion dynamic behavior of systems substituted with typical cations, which reveals the energetics for un-bounding the substituent cation–oxygen vacancy defect clusters. Thirdly, we examined the oxygen ion diffusion mechanisms to clarify the effects of size and charge of the substituent cations upon oxygen vacancy migration.

### 3.1 Defect energy

Fig. 1a represents the $CeO_2$ lattice with its centrosymmetric fluorite structure, where eight oxygen anions ($Ce^{4+}$-8O configuration) surround each cerium. Fig. 1b shows the local environment of $Ce^{4+}$ cations, whereas Fig. 1c presents the local structure of a substituent cation together with an oxygen vacancy at its nearest-neighboring site (7O–$\mathrm{Ca}_{\mathrm{Ce}}''$– $\mathrm{V}_{\mathrm{O}}^{\bullet\bullet}$configuration).

We first investigated the solution energy ($E_{sol}$) and binding energy ($E_{bind}$) values of various cations. Fig. 2a shows that the radii of the $M^{+}$ cations have little effect on the solution energies and binding energies. The considerable solution energies of the monovalent cations suggest a low solution limit of such cations in $CeO_2$, due to the significant charge difference between them and $Ce^{4+}$ cation. The similar size of $Na^{+}$ (1.02 Å) to $Ce^{4+}$ (0.98 Å)[43] results in the most favorable solution energy and the weakest interaction between $\mathrm{Na}_{\mathrm{Ce}}'''$ and $\mathrm{V}_{\mathrm{O}}^{\bullet\bullet}$ for the alkali metal cations. Likewise, there is a high correlation between the solution energy, the binding energy, and the size of the $M^{2+}$ cations (Fig. 2b). The reduced solution energies (ranging from 6.17 to 4.41 eV) suggest a higher solution limit than that for the monovalent cations. We expect that transition metals $Fe^{2+}$, $Co^{2+}$, $Mn^{2+}$, $Mg^{2+}$, $Ni^{2+}$, and $Zn^{2+}$ will act as significant trapping centers for oxygen vacancies due to their considerable binding energies (the magnitude of $E_{bind} > 2$ eV). The relatively low solution energies and the weak binding energies for $Ca^{2+}$ and $Sr^{2+}$ suggest that such cations are promising for the development of oxygen ion conductors as they will not exert a strong trapping effect on the oxygen vacancies.

For trivalent cations, size plays an essential role in determining the defect energy: the solution energy as well as the magnitude of the binding energy show a monotonic decrease with an increase of the ionic radius (Fig. 2c). The relatively large binding energies for $Al^{3+}$ and $Fe^{3+}$ indicate a strong interaction of $\mathrm{Al}_{\mathrm{Ce}}'$ and $\mathrm{Fe}_{\mathrm{Ce}}'$ with $\mathrm{V}_{\mathrm{O}}^{\bullet\bullet}$, which is attributed to the significant size mismatch with $Ce^{4+}$ in addition to the charge difference. Although there are limited experimental data about $\mathrm{Al}_{\mathrm{Ce}}'$ and $\mathrm{Fe}_{\mathrm{Ce}}'$ for direct comparison, Nuclear Magnetic Resonance spectroscopic studies on $^{45}Sc$ and $^{89}Y$ have revealed that all oxygen vacancies in the lattice are associated with $^{45}Sc$ or $^{89}Y$ atoms at the nearest neighboring site.[44, 45] The weak attractive interaction between large rare-earth cations (i.e., $Gd^{3+}$, $Eu^{3+}$, $Pr^{3+}$, and $Nd^{3+}$) and oxygen vacancies confirms that such cations are promising for the improvement of oxygen ion diffusivity, which is in line with previous experimental and theoretical studies.[9, 17] Unlike $M^+$, $M^{2+}$, and $M^{3+}$ cations, the incorporation of $M^{4+}$ cations will not introduce extrinsic oxygen vacancies. Therefore, such cations have rather low solution energies and are predicted to have a high solubility (Fig. 2d). This is consistent with the fact that $CeO_2$ can form solutions with $SiO_2$, $TiO_2$, and $ZrO_2$ over a wide composition range.[46, 47] The magnitude of the binding energy increases with an increase in the size mismatch, and the maximum value observed is for $Si^{4+}$. Such results demonstrate the importance of elastic strain effects, indicating that $M^{4+}$ cations prefer 7-fold coordination by oxygen ions, in contrast to the usual 8-fold coordination of $Ce^{4+}$ cations in $CeO_2$.

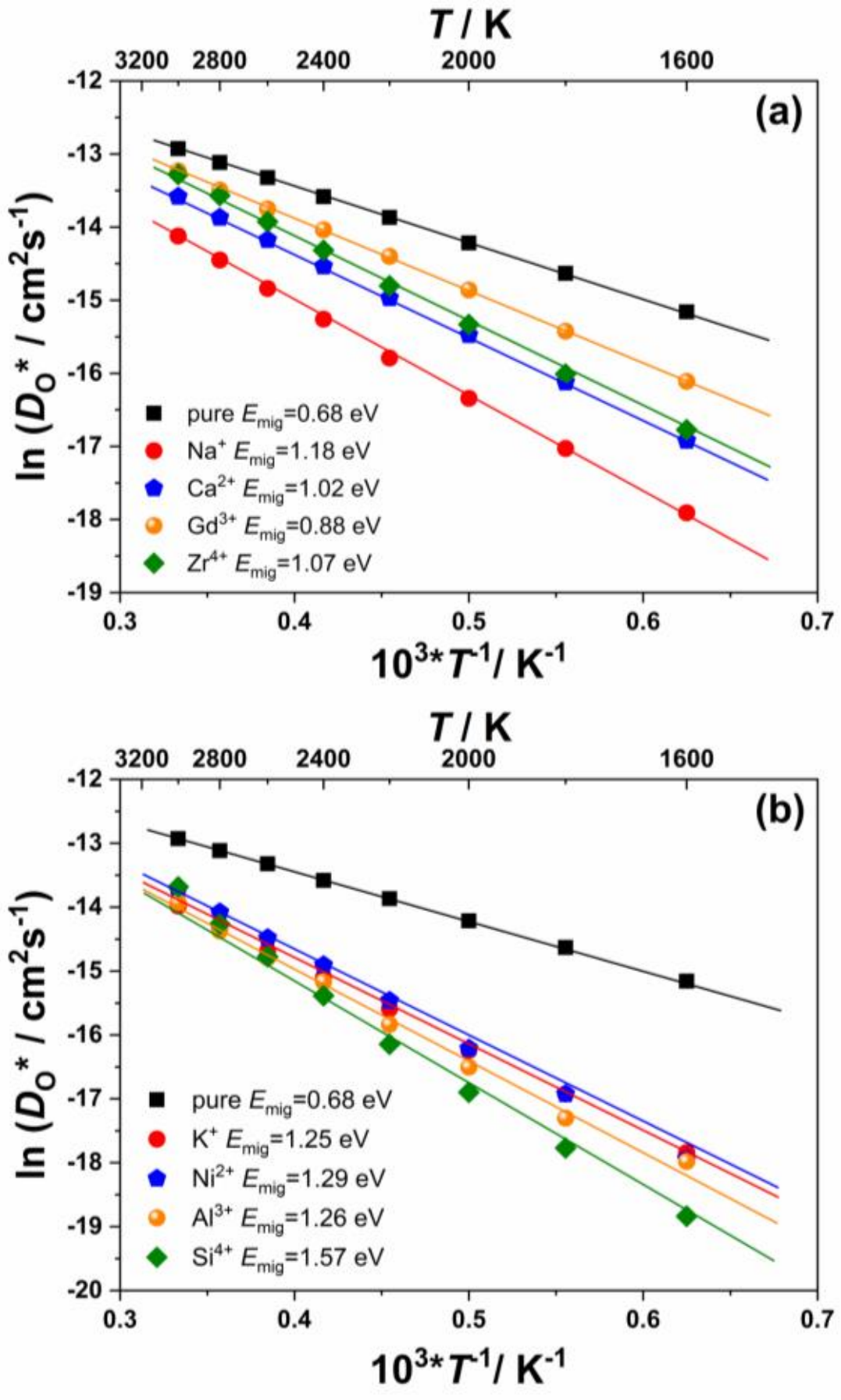

**Fig. 3.** Oxygen tracer diffusion coefficients ($D_O^*$) as a function of inverse temperature for systems substituted with (a) $Na^+$, $Ca^{2+}$, $Gd^{3+}$, and $Zr^{4+}$ and (b) $K^+$, $Ni^{2+}$, $Al^{3+}$, and $Si^{4+}$. The oxygen vacancy concentration is 1%.

### 3.2 Oxygen ion diffusion

The macroscopic oxygen ionic conductivity involves complex oxygen ionic-jump processes: the oxygen ion jumps around the defect center(s), the oxygen ion jumps away and towards the defect center(s), etc. We thus investigate the oxygen ion diffusion further by MD simulations with large supercells, which can take account of various possible jumps of the oxygen vacancies. We focus on typical cations, i.e., cations which show the weakest interactions (e.g. $Na^+$, $Ca^{2+}$, $Gd^{3+}$, and $Zr^{4+}$, 'weak set') and the strongest interactions (e.g., $K^+$, $Ni^{2+}$, $Al^{3+}$, and $Si^{4+}$, 'strong set') with oxygen vacancies within $M^+$, $M^{2+}$, $M^{3+}$, and $M^{4+}$ predicted by MS simulations. We do not consider $Li^+$ and $Rb^+$ due to their high mobility and low solution limit, respectively. Moreover, we select $Ca^{2+}$ rather than $Sr^{2+}$ for the following reasons: (1) $Ca^{2+}$ has a higher solution limit; (2) $Ca^{2+}$ (1.0 Å) has a similar size to $Na^+$ (1.02 Å) and $Gd^{3+}$ (0.94 Å);[43] and (3) because of the easy formation of the $SrCeO_3$ secondary phase within $Sr^{2+}$ cation-substituted $CeO_2$.[48]

Fig. 3a and b show the oxygen tracer diffusion coefficients ($D_O^*$) as a function of inverse temperature as well as the migration energies ($E_{mig}$) for oxygen vacancies obtained by MD simulations. We obtained the highest diffusion coefficient at any given temperature for the 'pure' system. This confirms the clustering of substituent cations with oxygen vacancies, which is consistent with the negative binding energies obtained by MS simulations (Fig. 2). Our results reveal that the migration energies of subvalent cation-substituted systems within the 'weak set' follow the order $Gd^{3+} > Ca^{2+} > Na^+$ (e.g., 0.88, 1.02 and 1.18 eV), whereas within the 'strong set' the migration energies are very close to each other (e.g., ≈1.30 eV). The migration energy increases significantly, from 0.68 eV for the 'pure' system to 1.07 eV for the $Zr^{4+}$-substituted system. We can also note that the $Si^{4+}$-substituted system exhibits the highest migration energy (1.57 eV) among all the systems investigated, as is consistent with the largest association energy between $\mathrm{Si}_{\mathrm{Ce}}^{\times}$ and the oxygen vacancy obtained by MS calculations (Fig. 2d).

Fig. 4a shows the oxygen tracer diffusion coefficient ($D_O^*$) as a function of the $Ca^{2+}$ concentration. With the increase of $Ca^{2+}$ concentration ($x$), the migration energy decreases slightly from 1.02 to 1.01 eV and then increases almost linearly to 1.27 eV for $x$ = 7.5%. Oxygen tracer diffusion coefficients as a function of oxygen vacancy concentration ($x_{V_O^{\bullet\bullet}}$) for $Gd_xCe_{1-x}O_{2-x/2}$, $Ca_xCe_{1-x}O_{2-x}$ and $Na_xCe_{1-x}O_{2-3x/2}$ are extrapolated to experimentally relevant temperatures and are summarized in Fig. 4b, c and d, respectively. For all the systems the values of $D_O^*$ show a typical increase and then a decrease with increasing $x_{V_O^{\bullet\bullet}}$. The decrease is attributed to the trapping of oxygen vacancies by defect centers. At 800 K, we can observe a $D_O^*$ maximum at $x_{V_O^{\bullet\bullet}}$ ≈ 2.5% for all systems investigated, corresponding to a concentration of $Gd^{3+}$, $Ca^{2+}$ and $Na^+$ being 10, 5 and 3.3%, respectively. The magnitude of $D_O^*$ varies by two orders of magnitude and follows the order $Gd_{0.1}Ce_{0.9}O_{1.95}$ > $Ca_{0.05}Ce_{0.95}O_{1.95}$ > $Na_{0.03}Ce_{0.97}O_{1.95}$. With increasing temperature, the optimum $x_{V_O^{\bullet\bullet}}$ slightly shifts to a higher value as overcoming the associating interactions becomes easier at higher temperatures.

Co-substituting two cation species is a promising approach for combining the positive effects of the substitution of single cations, that is beyond a mere additive effect.[18–20] Fig. S3 shows the oxygen ion diffusion behavior of the $Gd^{3+}/Ca^{2+}$, $Nd^{3+}/Ca^{2+}$ and $Lu^{3+}/Ca^{2+}$ co-substituted systems. We here fix the oxygen vacancy concentration ( $x_{V_O^{\bullet\bullet}}$ ) at 2.5% for ease of comparison. The $Gd^{3+}/Ca^{2+}$ and $Nd^{3+}/Ca^{2+}$ co-substitutions facilitate the oxygen ion diffusion and yield slightly higher tracer diffusion coefficients than the values obtained by a weighted average. By contrast, $Lu^{3+}$ and $Ca^{2+}$ co-substitution give rise to a significantly reduced $D_O^*$, indicating an enhanced association interactions between the defect clusters and the oxygen vacancies. Whereas such an association is detrimental for ionic conductivity it is desirable for electromechanical applications. Several experimental evidences indicate that oxygen ion blocking barriers can improve the electrostriction effect in ceria solutions.[12, 49]

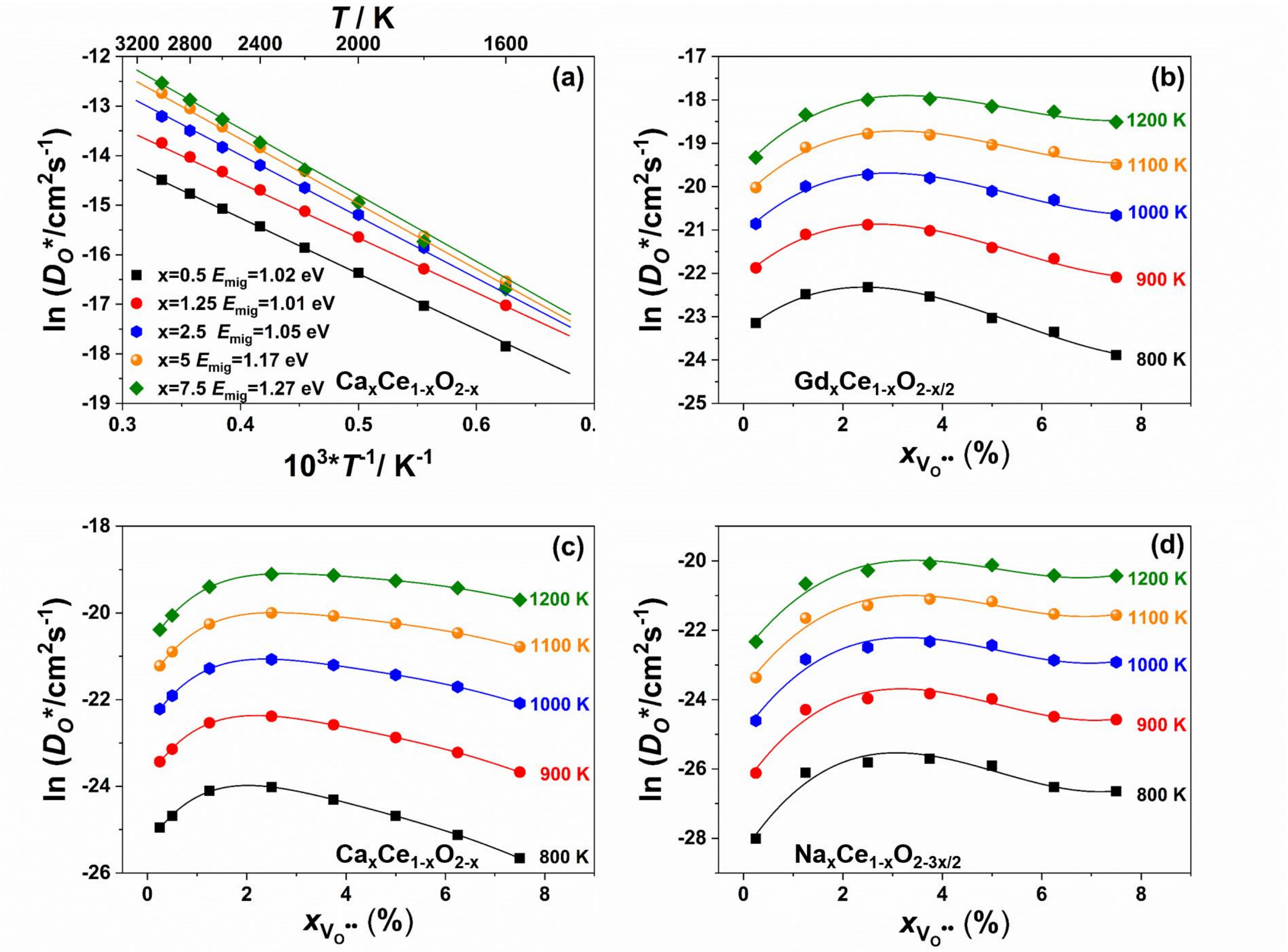

**Fig. 4.** (a) Oxygen tracer diffusion coefficients ($D_O^*$) as a function of inverse temperature for $CeO_2$ substituted by different $Ca^{2+}$ concentrations. Oxygen vacancy concentration ($x_{V_O^{••}}$ ) dependence of $D_O^*$ for (b) $Gd_xCe_{1-x}O_{2-x/2}$; (c) $Ca_xCe_{1-x}O_{2-x}$; (d) $Na_xCe_{1-x}O_{2-3x/2}$. $x$ in (a) corresponds to the $Ca^{2+}$ cation concentration.

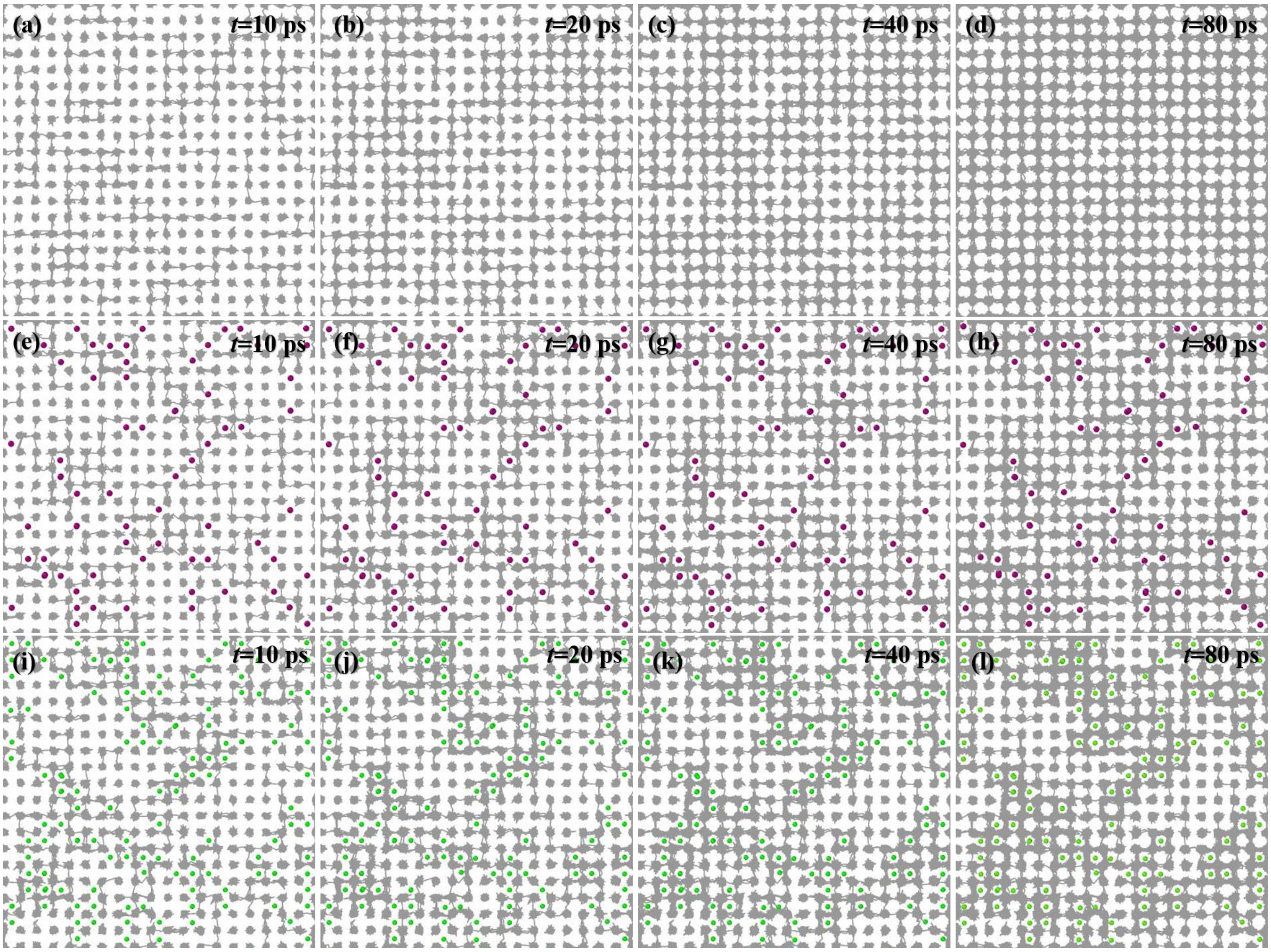

**Fig.** 5 Traced trajectories for oxygen ion diffusion of 'pure' system (a)–(d); $Ca^{2+}$ substituted system (e)–(h); $Si^{4+}$ substituted system (i)–(l) as a function of time at 2400 K projected onto the *ab* plane. The $Ce^{4+}$ and $O^{2-}$ ions are omitted for clarity. Note that the radii of the cations do not correspond to their real size. Colors: $Ca^{2+}$, purple; $Si^{4+}$, green.

**Table 1**. Energy barrier for blocking ($E_{block}$, eV) and trapping ($E_{trap}$, eV) mechanisms for oxide-ion migration with various diffusion pathways. The values for $E_{trap}$ correspond to an oxygen ion jumping from the nearest-neighboring site to the next nearest-neighboring site, whereas the values in the brackets correspond to the reverse process. Energy difference ($E_{diff}$, eV) of $E_{trap}$ compared to the oxygen ion migration within pure $CeO_2$. The effects of charge were examined by fixing the lattice configurations (oxygen ion migration with the presence of a $\mathrm{Ca}_{\mathrm{Ce}}''$ defect center), but replacing the $\mathrm{Ca}_{\mathrm{Ce}}''$ with $\mathrm{Na}_{\mathrm{Ce}}'''$ ($Na^+/Ca^{2+}$*), $\mathrm{Gd}_{\mathrm{Ce}}'$ ($Gd^{3+}/Ca^{2+}$*) or $\mathrm{Ce}_{\mathrm{Ce}}^{\times}$ ($Ce^{4+}/Ca^{2+}$*). In this approach, only the relaxation of the migrating oxygen ion is allowed to reduce the steric hindrance.

| | $Cd^{2+}$ | $Ca^{2+}$ | $Sr^{2+}$ | $Na^+$ | $Gd^{3+}$ | $Na^+/Ca^{2+}$* | $Gd^{3+}/Ca^{2+}$* | $Ce^{4+}/Ca^{2+}$* |
|---|---|---|---|---|---|---|---|---|
| $E_{block}$ | 0.67 | 0.75 | 1.09 | 0.88 | 0.65 | 0.60 | 1.02 | 0.78 |
| $E_{trap}$ | 0.16(0.78) | 0.27(0.58) | 0.35(0.54) | 0.18(0.72) | 0.53(0.66) | 0.13(0.68) | 0.37(0.44) | 0.75(0.76) |
| $E_{diff}$ | −0.37(0.25) | −0.26(0.05) | −0.18(0.10) | −0.35(0.19) | 0(0.13) | −0.40(0.15) | −0.16(-0.09) | 0.22(0.23) |

*Oxygen ion diffusion with a fixed initial-, final- and saddle-point configuration of $\mathrm{Ca}_{\mathrm{Ce}}''$

### 3.3 Traced trajectories

To better visualize the diffusion pathways and all positions in the lattice traversed as a function of time, we produce the traced trajectories for oxygen ion diffusion as a function of time at $T$ = 2400 K (Fig. 5). As expected, the 'pure' system exhibits significant long-range, three-dimensional oxygen ion diffusion. The isotropic and homogeneous traced trajectories demonstrate that the oxygen ion follows a three-dimensional random walk behavior. From the traced migration paths shown in Fig. 5e–l, it is evident that the substituting defects will act as trapping centers for oxygen vacancies, making the oxygen vacancies 'rattle around' such defect centers. This effect certainly reduces the number of 'free' oxygen vacancies available for migration. Therefore, the long-range diffusion of oxygen ions is significantly restricted and the $D_{\mathrm{O}}^{*}$ value decreases in aliovalent/isovalent cation-substituted $CeO_2$ systems. It is worth mentioning here that the oxygen vacancies in our systems are generated by randomly removing the oxygen ions from the lattice sites. The traced trajectories for $Ca^{2+}$ cation substituted materials show evident enrichment around such cation defects as a function of time. Therefore, the substituting defect centers will not only affect the energy barriers for oxygen ion diffusion but will also affect the distribution of oxygen vacancies within the crystal lattice. This configuration is also the case for $Si^{4+}$ cations, but with a much more profound effect (Fig. 5c). A comparison of the traced trajectories for $Na^{+}$ cation-substituted systems and $K^{+}$-substituted ones reveals that (see Fig. S4) the $K^{+}$ cations would reduce the channels for the smooth movement of oxygen ions around the $K'''_{Ce}$ defect centers, which is related to their larger ionic size.

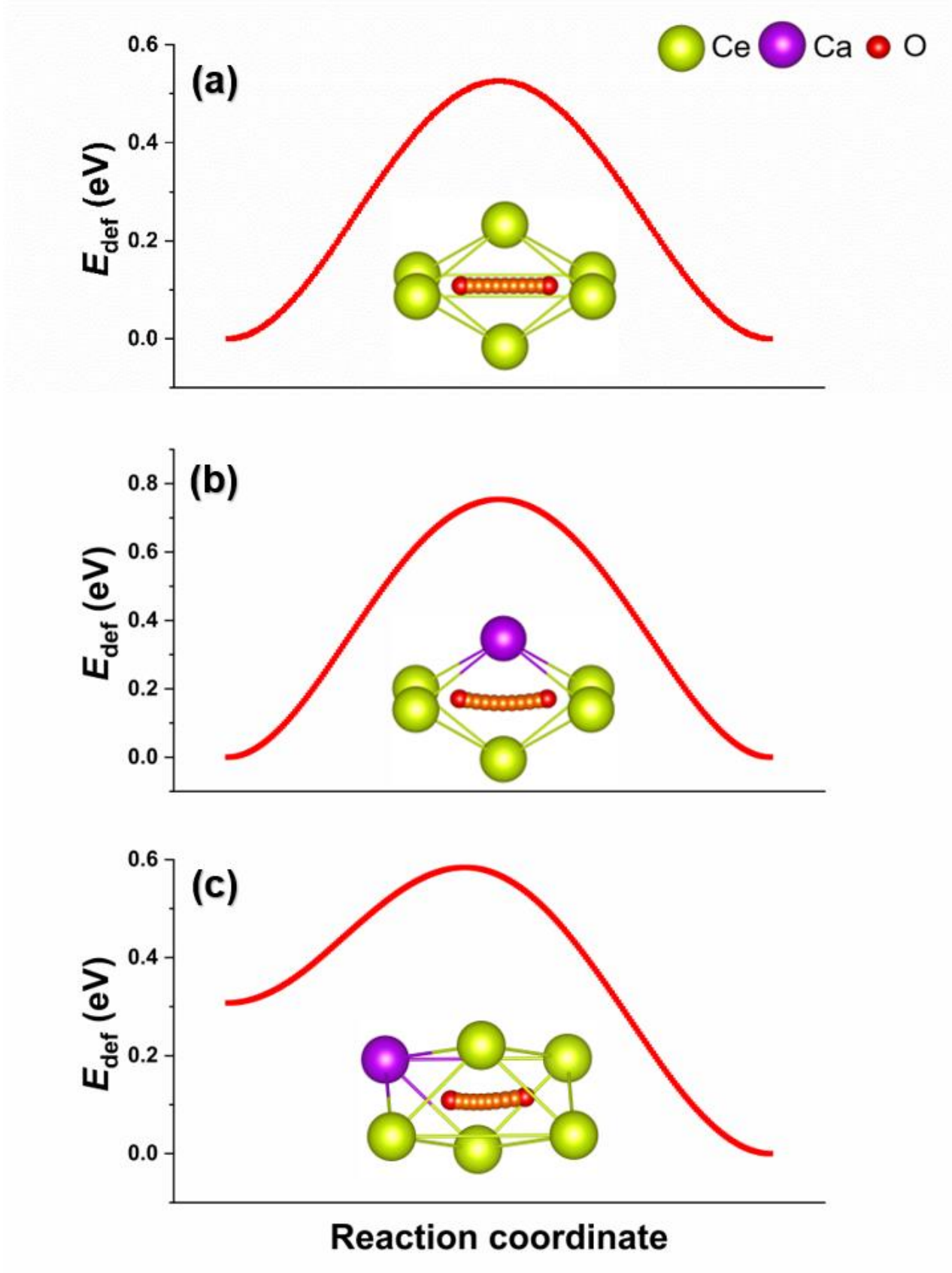

**Fig. 6** Defect site energy as a function of the reaction coordinate for oxygen ion migration with various migration edges: (a) within pure $CeO_2$; (b) from one nearest-neighboring site to another nearest-neighboring site of $Ca''_{Ce}$; (c) from the nearest-neighboring site to the next nearest-neighboring site of $Ca''_{Ce}$ and *vice versa*. The insets are the traced trajectories, which show that the oxygen ion follows a linear pathway in (a), but a slightly curved pathway in (b) and (c). The energy profile for (c) corresponds to an oxygen ion migrating from the left side to the right side (with the oxygen vacancy direction from the right side to the left side). Colors: $Ce^{4+}$, yellow; $Ca^{2+}$ cation, purple; oxygen, red.

### 3.4 Oxygen ion migration energy

As shown in Fig. 6a, the most favourable pathway within pure $CeO_2$ for an oxygen ion to migrate is linearly along the (100) direction, with an energy barrier of 0.53 eV. This agrees well with the results obtained by MS simulations (0.59 eV, Fig. S1). Fig. 6b and c exhibit the 'blocking mechanism' ($E_{block}$) and 'trapping mechanism' ($E_{trap}$),[50, 51] respectively, of $Ca''_{Ce}$ on the migrating oxygen ion, i.e., the oxygen ion migrates from one nearest-neighboring site to another nearest-neighboring site (blocking); and from the nearest-neighboring site to the next nearest-neighboring site and *vice versa* (trapping). For the former migration mechanism (Fig. 6b), the presence of $Ca''_{Ce}$ yields a slightly curved pathway with the saddle-point deviating away from $Ca''_{Ce}$, and an increase of the energy barrier to 0.75 eV. For the latter migration mechanism (Fig. 6c), the trapping of oxygen vacancy by $Ca''_{Ce}$ results in an asymmetric energy profile, with the migration energy for the vacancy jump away from the associating $Ca''_{Ce}$ (0.58 eV) being much higher than for the reverse process (0.27 eV). This is consistent with the high binding energy obtained by MS lattice simulations and confirms that the oxygen vacancy prefers to reside in the nearest-neighboring site of the $Ca''_{Ce}$.

The effects of the substituent cation's size are analyzed in detail by inspecting the oxygen ion migration along various migration edges, as listed in Table 1. Compared with $Ca''_{Ce}$, the $Cd''_{Ce}$ defect center increases the local free volume, which facilitates the oxygen ion migration and thereby reduces the $E_{block}$ to 0.67 eV. By contrast, the $Sr''_{Ce}$ defect center reduces the free volume for the oxygen ion to jump through the migration aperture. Therefore, the constituent ions should relax substantially to reduce the steric hindrance to allow the jump to take place, thus increasing the energy barrier to 1.09 eV. On the other hand, $E_{trap}$ decreases with an increase of the subvalent cation radius, as in agreement with the results obtained by MS simulations. To identify the effects of charge, we fix the lattice configurations (oxygen ion migration with the presence of a $Ca''_{Ce}$ defect center), but replace $Ca''_{Ce}$ by $Na'''_{Ce}$, $Gd'_{Ce}$ or $Ce^{\times}_{Ce}$ (abbreviated as $Na^{+}/Ca^{2+}$*, $Gd^{3+}/Ca^{2+}$*, and $Ce^{4+}/Ca^{2+}$*, respectively). In this way, only the relaxation of the migrating oxygen ion is allowed to reduce the steric hindrance. The increased $E_{trap}$ indicates that the interactions between the migrating oxygen ion and the defect center are becoming more and more difficult to overcome as the subvalent cation charge increases. Moreover, the values of $E_{block}$ follow the trend $Gd^{3+}/Ca^{2+}$* > $Gd^{3+}$ > $Na^{+}$ > $Na^{+}/Ca^{2+}$*. Therefore, extensive lattice relaxations are involved in the realization of oxygen ion migration, the magnitude of which follows the order $Gd^{3+}$ > $Ca^{2+}$ > $Na^{+}$.

Owning to the significant lattice distortion, the $Ce^{4+}/Ca^{2+}$* system results in a lower nearest-neighboring site energy than the next nearest-neighboring site for oxygen ions, indicating that the oxygen vacancy prefers to locate nearby the $Ce^{4+}/Ca^{2+}$* defect center. Furthermore, the relatively high blocking energy values for the $K^{+}$ (1.26 eV) and $Ba^{2+}$ (1.50 eV) cations (Table S3) further demonstrate that this migration is highly unfavorable, which is consistent with traced trajectories observed by MD simulations.

## 4. Conclusions

Based on atomic-scale simulations calculations, we identify useful trends in the defect chemistry and the oxide ion diffusion mechanism in $CeO_2$-based material systems:

1. For subvalent cations of similar size, the higher the charge value, the higher the oxygen diffusivity (i.e. the trend $Gd^{3+}$ > $Ca^{2+}$>

Na$^+$ is followed) and the weaker the associating interaction between the oxygen vacancies and the dopants.
2. For isovalent cations, the size mismatch to $Ce^{4+}$ yields a higher oxygen ionic diffusivity, e.g., the trend $Na^+ > K^+$, $Ca^{2+} > Ni^{2+}$, $Gd^{3+} > Al^{3+}$ is followed. More specifically, larger cations block the oxygen ion migration because they reduce the 'free volume' for the movement of the oxygen ions. On the other hand, smaller cations lower the energy barrier for the oxygen vacancy to 'rattle around', which reduces the number of 'free' oxygen vacancies available for ionic conduction. In addition, the magnitude of the binding energies increases with increase in the size mismatch, highlighting the importance of elastic strain effects.
3. To achieve fast oxygen ionic transport the optimum oxygen vacancy concentration is 2.5% for $Gd_xCe_{1-x}O_{2-x/2}$, $Ca_xCe_{1-x}O_{2-x}$ and $Na_xCe_{1-x}O_{2-3x/2}$ at 800 K, which is not constant and shifts gradually to higher values with increasing temperature.
4. Co-substitutions can enhance the impact of single substitutions beyond that expected by simple addition.

In addition to identifying the optimal dopants for new materials with improved properties, these fundamental insights into defect chemistry, defect–defect interactions, and oxygen ion diffusion dynamics and mechanisms can be used to accelerate the design of ceria defect systems.

## Conflicts of interest

There are no conflicts of interest to declare

## Acknowledgements

The authors acknowledge funding by The Danish Council for Independent Research Technology and Production Sciences for the DFF-Research Project 2, grant no. 48293 (Giant-E) and 6111-00145B (NICE); the European Union's Horizon 2020 research and innovation programme under Grant Agreement No. 801267 (BioWings); and VILLUM FONDEN, grant no. 00022862 (Iride).

## Supplementary Information

# Atomic-scale insights into electro-steric substitutional chemistry of cerium oxide

Haiwu Zhang, Ivano E. Castelli*, Simone Santucci, Simone Sanna, Nini Pryds, and Vincenzo Esposito*

Department of Energy Conversion and Storage, Technical University of Denmark, Anker Engelunds Vej 411, DK-2800 Kgs. Lyngby, Denmark

*Corresponding authors: Ivano E. Castelli (ivca@dtu.dk), Vincenzo Esposito (vies@dtu.dk)

## 1. Potential parameters for classical simulations.

**Table S1**. Interatomic potential parameters for CeO2. The Cutoff energy is 15 Å.

| Buckingham parameters | | | | Shell model parameters | |
|---|---|---|---|---|---|
| M-$O^{2-}$ | *A*/eV | *ρ*/Å | *C*/eV Å$^6$ | *Y*/e | *K*/eV Å$^{-2}$ |
| $Ce^{4+}$-$O^{2-}$ | 1986.83 | 0.3511 | 20.40 | -3.7 | 291.75 |
| $O^{2-}$-$O^{2-}$ | 22764.3 | 0.149 | 27.88 | -2.67 | 74.92 |

The potential parameters for $Ce^{4+}$-$O^{2-}$ and $O^{2-}$-$O^{2-}$ were directly taken from previous work on reduction and oxygen migration in ceria based oxides by G. Balducci *et al.*[1] The potential parameters for $Hf^{4+}$-$O^{2-}$ and $Zr^{4+}$-$O^{2-}$ were also taken from this work.[1] Potential parameters for $Li^+$, $K^+$, $Rb^+$, $Fe^{2+}$, $Co^{2+}$, $Zn^{2+}$, $Ca^{2+}$, $Sr^{2+}$, $Ba^{2+}$, $Al^{3+}$, $Sc^{3+}$, $Si^{4+}$, $Ge^{4+}$ and $Sn^{4+}$ were taken from J. R. Tolchard and M. S. Islam's previous work on doping effects in apatite silicate ionic conductors.[2] For $Na^+$ and $Ti^{4+}$, the potential parameters were taken from atomistic simulation work on sodium bismuth titanate by H. Zhang *et al.*[3] For $Cd^{2+}$, $Gd^{3+}$, $Nd^{3+}$, $Y^{3+}$, $Ni^{2+}$, $Yb^{3+}$ and $Lu^{3+}$, the potential parameters were taken from ref 4 by G. V. Lewis and C. R. A. Catlow. Potential parameters for other cations were taken from other related works: $Mg^{2+}$, $In^{3+}$ and $La^{3+}$ from ref 5; $Mn^{2+}$ from ref 6; $Fe^{3+}$ from ref 7; $Eu^{3+}$ and $Pr^{3+}$ from ref 8.

## 2. Lattice parameters

**Table S2**. Comparison of experimental lattice constant (*a*, Å) and bond of Ce-O ($d_{Ce\text{-}O}$, Å) of pure $CeO_2$ with simulated results.

| | Exp | Classical simulations | LDA | GGA | GGA+U |
|---|---|---|---|---|---|
| *a*, Å | 5.412,[9] 5.407[10] | 5.429 | 5.546 | 5.406 | 5.445 |
| $d_{Ce\text{-}O}$, Å | 2.34346,[9] 2.3413[10] | 2.35096 | 2.40149 | 2.34079 | 2.3581 |

## 3. Full list of defect equations

1) Monovalemt ($M^+$) cations: $\mathrm{M_2O + 2Ce_{Ce}^{\times} + 3O_O^{\times} \rightarrow 2M_{Ce}^{\prime\prime\prime} + 3V_O^{\bullet\bullet} + 2CeO_2}$

$$E_{sol,mono} = 2E_{CeO_2} + 3E_{def}(\mathrm{V_O^{\bullet\bullet}}) + 2E_{def}(\mathrm{M_{Ce}^{\prime\prime\prime}}) - E_{M_2O}$$

2) Divalent ($M^{2+}$) cations: $\mathrm{MO + Ce_{Ce}^{\times} + O_O^{\times} \rightarrow M_{Ce}^{\prime\prime} + V_O^{\bullet\bullet} + CeO_2}$

$$E_{sol,div} = E_{CeO_2} + E_{def}(\mathrm{V_O^{\bullet\bullet}}) + E_{def}(\mathrm{M_{Ce}^{\prime\prime}}) - E_{MO}$$

3) Trivalent ($M^{3+}$) cations: $\mathrm{M_2O_3 + 2Ce_{Ce}^{\times} + O_O^{\times} \rightarrow 2M_{Ce}^{\prime} + V_O^{\bullet\bullet} + 2CeO_2}$

$$E_{sol,tri} = 2E_{CeO_2} + E_{def}(\mathrm{V_O^{\bullet\bullet}}) + 2E_{def}(\mathrm{M_{Ce}^{\prime}}) - E_{M_2O_3}$$

4) Tetravalent ($M^{4+}$) cations: $\mathrm{MO_2 + Ce_{Ce}^{\times} \rightarrow M_{Ce}^{\times} + CeO_2}$

$$E_{sol,tetr} = E_{CeO_2} + E_{def}(\mathrm{M_{Ce}^{\times}}) - E_{MO_2}$$

## 4. Energy profile obtained by classical simulations

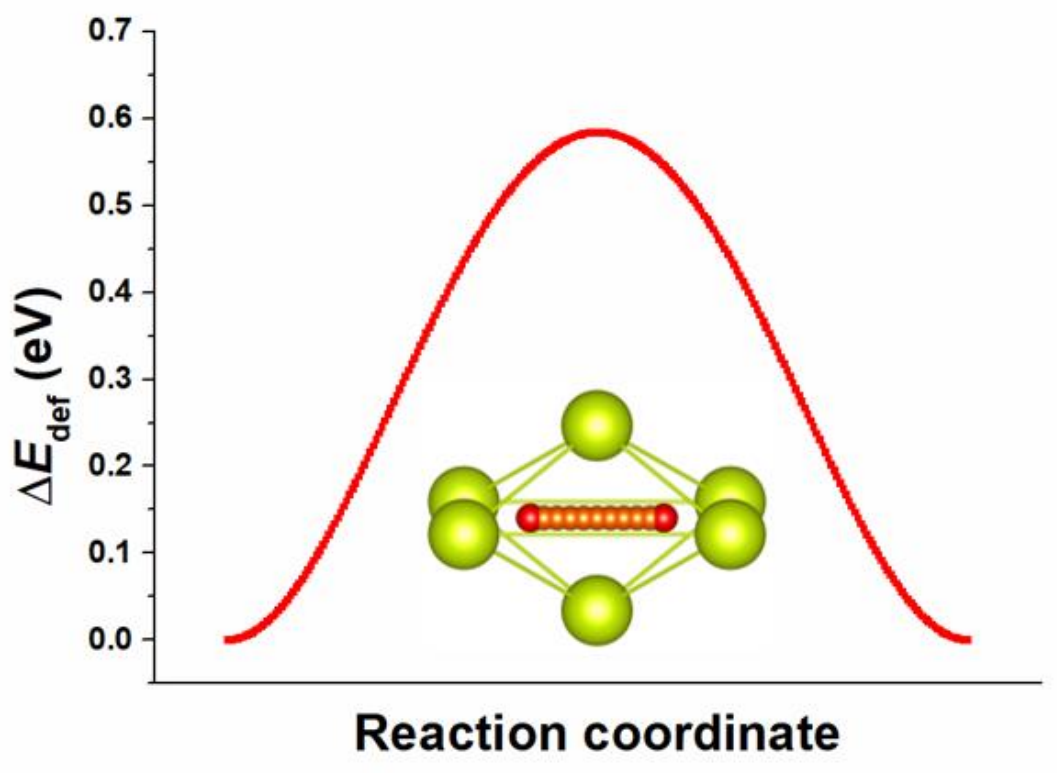

**Fig. S1** Defect site energy as a function of reaction coordinate for oxygen ion migration within pure $CeO_2$ obtained by MS simulations using GULP.

## 5. Mean-squared displacements for 'pure' system

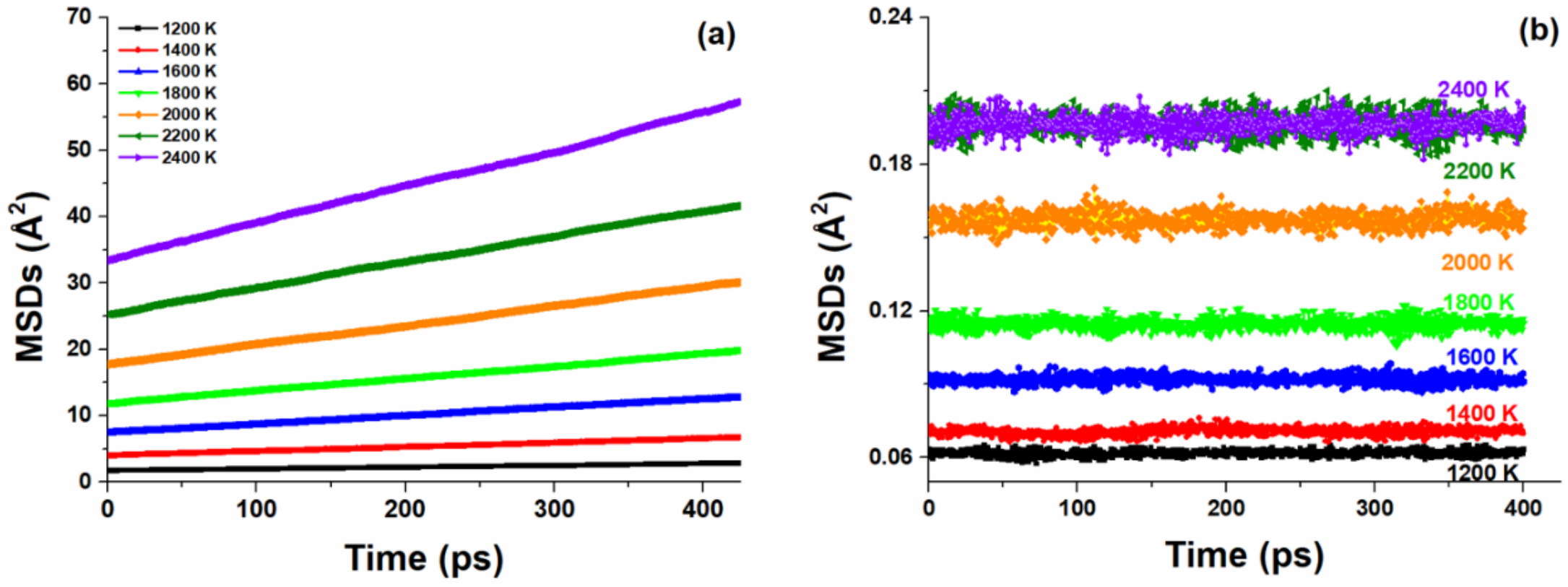

**Fig. S2**. Temperature dependent MSDs of (a) oxygen ions and (b) $Ce^{4+}$ cations of 'pure' $CeO_2$.

## 6. Oxygen ion diffusion in co-substituted systems

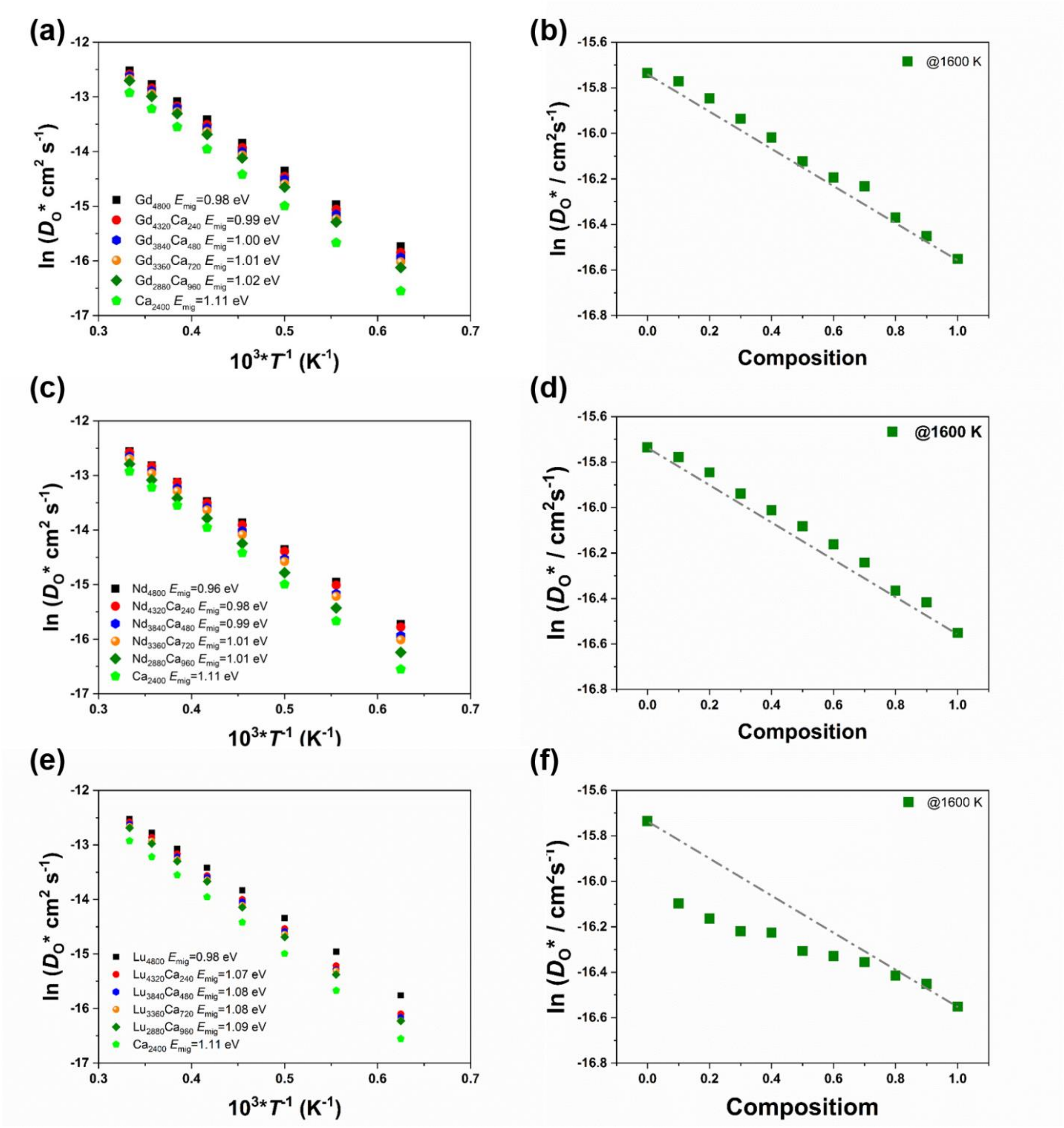

**Fig. S3**. Inverse temperature dependent oxygen tracer diffusion coefficients ($D_O^*$) for (a) $Gd^{3+}$/$Ca^{2+}$; (c) $Nd^{3+}$/$Ca^{2+}$; (e) $Lu^{3+}$/$Ca^{2+}$ co-substituted systems. Oxygen tracer diffusion coefficients ($D_O^*$) for (b) $Gd^{3+}$/$Ca^{2+}$; (d) $Nd^{3+}$/$Ca^{2+}$; (f) $Lu^{3+}$/$Ca^{2+}$ co-substituted systems as a function of $Ca^{2+}$ concentration. The lines in (b), (d) and (f) represent the expected $D_O^*$ based on a weighted average. The oxygen vacancy concentration ($x_{V_O^{\bullet\bullet}}$ ) is 2.5% for all the co-substituted systems.

## 7. Traced trajectories for various cations substituted systems

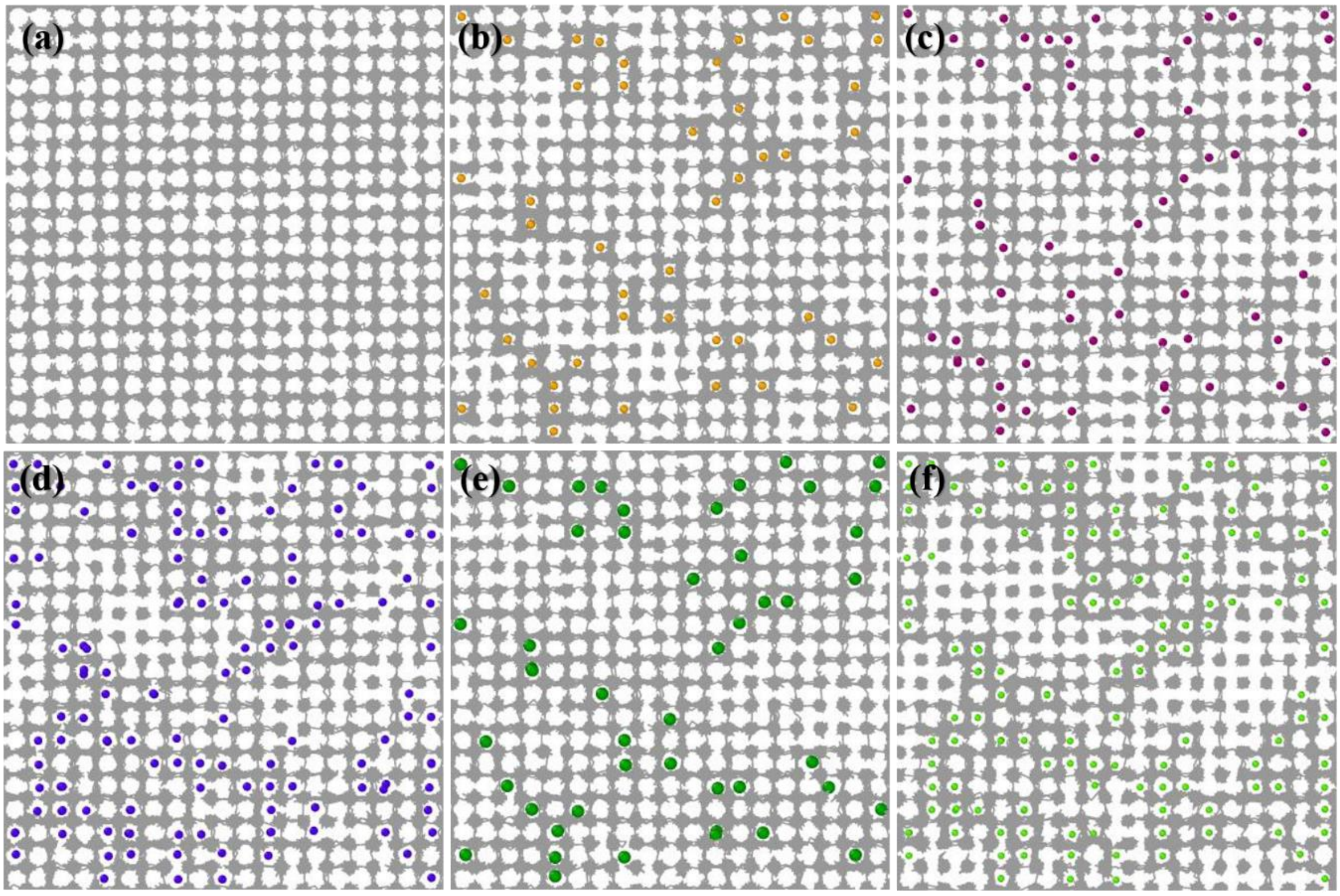

**Fig. S4**. Traced trajectories for oxygen ion of (a) pure $CeO_2$; and $CeO_2$ substituted by (b) $Na^+$; (c) $Ca^{2+}$; (d) $Gd^{3+}$; (e) $K^+$, and (f) $Si^{4+}$ at *t*=80 ps at 2400 K projected onto the *ab* plane. The $Ce^{4+}$ and $O^{2-}$ ions are omitted for clarify. Note that the radii of the cations does not correspond to the real size. Key: Na: orange, Ca: purple, Gd: violet, K: olive, Si: green.

## 8. Oxygen ion migration

**Table S3**. Energy barrier for blocking ($E_{block}$, eV) and trapping ($E_{trap}$, eV) mechanisms for oxygen ion migration with various diffusion pathways. The values for $E_{trap}$ correspond to an oxygen ion to jump from the nearest-neighbouring site to the next nearest-neighbouring site, whilst the values in the brackets correspond to the reverse process. Energy difference ($E_{diff}$, eV) of $E_{trap}$ with respect to oxygen ion to migrate within pure $CeO_2$.

| | $K^+$ | $Fe^{2+}$ | $Ba^{2+}$ | $Al^{3+}$ | $K^+/Ca^{2+}$* | $Fe^{2+}/Ca^{2+}$* | $Cd^{2+}/Ca^{2+}$* | $Sr^{2+}/Ca^{2+}$* |
|---|---|---|---|---|---|---|---|---|
| $E_{block}$ | 1.32 | 0.22 | 1.50 | 0.51 | 1.18 | 0.11 | 0.60 | 1.02 |
| $E_{trap}$ | 0.13(0.58) | 0.0(0.76) | 0.0(0.18) | 0.60(1.18) | 0.30(0.53) | 0.14(0.96) | 0.13(0.68) | 0.37(0.44) |
| $E_{diff}$ | -0.43(0.02) | -0.56(0.20) | -0.56(-0.38) | 0.04(-0.62) | -0.26(-0.03) | -0.42(0.40) | -0.43(0.12) | -0.19(-0.12) |

*oxygen ion diffusion with a fixed saddle point configuration of $Ca''_{Ce}$